\documentclass{svjour3}                     
\smartqed  

\usepackage[dvips]{psfrag, graphicx, epsfig}
\def\bold1{\boldsymbol{1}}
\def\bold0{\boldsymbol{0}}

\begin{document}

\title{The time to extinction for an SIS-household-epidemic model}

\titlerunning{Time to extinction for an SIS-household epidemic}        

\author{Tom Britton        \and
        Peter Neal 
}


\institute{T. Britton \at
               Department of
Mathematics, Stockholm University, SE-106 91 Stockholm, Sweden. \\
\email{tom.britton@math.su.se}
           \and
           P.Neal \at
             School of Mathematics, Alan Turing Building, University
of Manchester, Manchester, M13 9PL, UK. \\Tel.: +44-161-3063634 \\
              Fax: +44-161-3063669\\
\email{peter.neal@manchester.ac.uk}  }

\date{Received: 18 June 2009 / Accepted: date}

\maketitle

\begin{abstract}
We analyse a stochastic SIS epidemic amongst a finite population
partitioned into households. Since the population is finite, the
epidemic will eventually go extinct, {\it i.e.}~ have no more
infectives in the population. We study the effects of population
size and within household transmission upon the time to extinction.
This is done through two approximations. The first approximation is
suitable for all levels of within household transmission and is
based upon an Ornstein-Uhlenbeck process approximation for the
diseases fluctuations about an endemic level relying on a large
population. The second approximation is suitable for high levels of
within household transmission and approximates the number of
infectious households by a simple homogeneously mixing SIS model
with the households replaced by individuals. The analysis, supported
by a simulation study, shows that the mean time to extinction is
minimized by moderate levels of within household transmission.
\end{abstract}

\section{Introduction}
Epidemic models are widely used for increasing the understanding of
infectious disease dynamics ({\it e.g.}~Anderson and May
\cite{AM91}) and for determining suitable preventive measures to
reduce or ideally stop disease spread ({\it e.g.}~Wallinga et al.
\cite{WT04}, Ferguson et al. \cite{Ferg06} and Cauchemez et al.
\cite{Cau2008}). In the present work we study a class of models for
diseases that are endemic. More precisely we look at SIS-epidemic
models (Kryscio and Lef\`{e}vre \cite{KL89}) meaning that
individuals are either susceptible (S) or infectious (I), and
susceptible individuals might get infected and infectious, and after
a time in the infectious state individuals recover and return to the
susceptible state. The community is considered closed being of size
$N$, but an alternative interpretation of the model is that an
individual that stops being infectious becomes immune for the rest
of its life (or dies) but is ``replaced'' by a new susceptible
individual when the infectious period stops.

SIS-epidemic models can potentially lead to the disease establishing
itself within the population and becoming endemic for a long period
of time. However, eventually the disease goes extinct. It is
therefore important to study properties of the model that determine
whether the disease may become endemic, and if so what the endemic
level is and for how long (on average) it will take for the disease
to go extinct. In terms of prevention the answers to these questions
can give guidance in determining how effective a given preventive
measure is, for example: what proportion is necessary to vaccinate
in order to make the disease go extinct promptly?

In the present paper we address the question of the time until an
endemic disease goes extinct, where we assume the community size $N$
is large and that we start in the endemic level. More precisely we
study the situation where individuals are grouped into households
and where it is assumed that transmission is much higher between
individuals in the same household as compared with individuals in
different households. We derive an approximation for the expected
time to extinction and compare it with the case that there are no
households (homogeneous mixing) for which asymptotic results for the
time to extinction are already available (Andersson and Djehiche
\cite{AD98}). This paper is hence concerned with analysing how the
time to extinction is affected when the assumption of homogeneous
mixing of identical individuals is relaxed by admitting the presence
of households. Similar questions have been analysed when relaxing
the assumption of homogeneous mixing but in other "directions":
Hagenars et al. \cite{Hag04} look at the case with an SIR epidemic
where the community consists of a few large sub-communities assuming
higher contact rates within the sub-communities, Lindholm
\cite{Lind08} studies an epidemic model where he investigates how
the time to extinction is affected by individuals being of different
types having different susceptibility and infectivity to the
disease.

Our main conclusion is that, assuming a fixed endemic level of
infectives $\hat i$, the time to extinction is smaller if there is
moderate or high transmission within households as compared to the
case of homogeneous mixing in the limit as $N \to \infty$. For high
values of $\hat i$ our Ornstein-Uhlenbeck approximation suggests a
monotone decrease of the expected time to extinction as a function
of the degree of within-household transmission rate $\lambda_L$,
whereas for small $\hat i$-values the expected time to extinction
first increases and then decreases as a function of $\lambda_L$. In
the latter case the longest time to extinction is hence for the case
with small (but positive!) $\lambda_L$. The Ornstein-Uhlenbeck
approximation is an asymptotic result as $N\to\infty$ using a normal
approximation for the probability of being close to extinction. We
introduce a second approximation which is appropriate when the
within-household transmission rate, $\lambda_L$, tends to infinity.
This approximation shows that for fixed population size $N$, the
mean time to extinction grows polynomially with $\lambda_L$ as
$\lambda_L \to \infty$ with the rate of the polynomial growth
depending upon the maximum household size. Thus there are two
competing approximations leading in practice to the mean time to
extinction being smallest for moderate levels of within household
infection. We focus the analytical analysis on the case with
households of size 2 with a simulation study showing that
qualitatively similar results hold for households having other and
possible different household sizes.

In Section 2 we define the SIS-household epidemic model and present
some known results for the model (Ball \cite{Ball99}, Neal
\cite{Neal06}). In Section 3 we derive our two approximations for
the expected time to extinction and give a full discussion of the
merits of the two approximations. In Section 4 we give some
numerical examples and plot the expected time to disease extinction
as a function of the degree of transmission within households, and
compare these approximations with simulation results. The paper ends
with a short discussion about the limitations of the present
analysis, and a discussion about interesting related
models/questions.

\section{The SIS household epidemic model}
\subsection{Model definition}\label{SIS-def}

We now define the Markovian household SIS epidemic model. This a
special case of the more general household SIS epidemic model
defined and analysed by Ball \cite{Ball99}.

Consider a fixed community consisting of $n$ households, all being
of the same size $k$ for simplicity of exposition (later we will
primarily focus on the case $k=2$). The community size is hence
$N=kn$, and we assume that $n$ and hence $N$ is large. Each
individual is at any time either infectious or susceptible.
Susceptible individuals recover (and become susceptible again)
independently at the constant rate $\gamma$. While infectious an
individual makes two types of contacts: the individual has
\emph{global} ``close contacts'' at rate $\lambda_G$, each time the
contacted person is selected uniformly at random from the whole
community, and the individual has \emph{local} ``close contacts'' at
rate $\lambda_L$ with individuals belonging to the same household,
here too the individual to be contacted is selected uniformly at
random, but now among the individuals belonging to the same
household as the person in question. By ``close contact'' we mean a
contact that results in infection if the other person is susceptible
-- if the other person is infectious the contact is assumed to have
no effect. The epidemic is initiated by some given initial condition
specifying how many households have $j$ infective individuals, $j=0,
1, \ldots, k$ infective individuals.

Let $Y_j(t)$ denote the number of households having $j$ infectives
(and hence $k-j$ susceptibles) at $t$. The initial condition is
hence specified by the vector $(Y_0(0),\dots , Y_k(0))$, for any
$t$: $\sum_jY_j(t)=n$ since the total number of households is $n$.
Let $I(t)=\sum_jjY_j(t)$ denote the number of infectives
(=infectious individual) at $t$, and similarly
$S(t)=\sum_j(k-j)Y_j(t)$ is the number of susceptible individuals.

The process $(Y_0(t),\dots , Y_k(t))$ is Markovian and there is at
most one infection or recovery occurring at a given time point,
implying the possible jumps are that one component decreases by 1
and at the same time the component directly to the left (recovery)
or directly to the right (infection) increases by 1. From the model,
each individual recovers at rate $\gamma$, so the rate at which some
household having $i$ infectives at present has a recovery (implying
that it changes to state $i-1$) equals $\gamma \cdot iY_i(t)$. An
infection can occur in two different ways. Either an individual is
infected globally, or else locally from within the household. The
rate of being infected globally is the same for all susceptible
individuals: $\lambda_GI(t)/N$, each infective has close contact
with a given individual at rate $\lambda_G/N$ and there are $I(t)$
infectives (in principle the division should be with $N-1$ since it
is not possible to have contact with yourself, but when $N$ is large
this is negligible). Additionally, a susceptible in a household
having $i$ infectives at present gets infected locally at rate
$i\lambda_L/(k-1)$ since each infective has close contact with a
specific household member at rate $\lambda_L/(k-1)$. The overall
rate at which some household having $i$ infectives moves to the
state $i+1$ is hence
\[
\left( \lambda_GI(t)/N + i\lambda_L/(k-1)\right)(k-i).
\]

\subsection{Properties of the SIS household epidemic model}\label{SIS-prop}

From now on we restrict ourselves to the case where all households
are of size $k=2$. The reason for this is to simplify notation and
keeping the dimensions of vectors and matrices low -- there is
nothing harder in principle for larger household sizes. The SIS
household epidemic model have been analysed previously be Ball
\cite{Ball99}, Ghoshal {\it et al.}~\cite{Ghos04} and Neal
\cite{Neal06}, in particular studying how the model behaves when $n$
is large. To this end we define $\bar Y_i (t)=Y_i(t)/n$, the
proportion of households having $i$ infectives and let $\bar{I} (t)
= I (t)/N$, the proportion of the population infectious. In Ball
\cite{Ball99} it is shown that, as $n\to\infty $, the process $(\bar
Y_0(t), \bar Y_1(t) , \bar Y_2(t))$ converges in probability
(uniformly on finite intervals) to the deterministic vector process
$(y_0(t), y_1(t) , y_2(t))$ defined as the solution to the
differential equations
\begin{eqnarray*}
y_0'(t)&= \gamma y_1(t)-\lambda_Gi(t)2y_0(t)
\\
y_1'(t)&= \lambda_Gi(t)2y_0(t) + 2\gamma y_2(t) -\gamma y_1(t) -
(\lambda_Gi(t)+\lambda_L)y_1(t)\label{det-diff}
\\
y_2'(t)&= - 2\gamma y_2(t) + (\lambda_Gi(t)+\lambda_L)y_1(t),
\end{eqnarray*}
where $i(t)=\sum_iiy_i(t)/2$ (the fraction of infectives). The
result relies on the initial conditions agreeing and having a
positive fraction of infectives: $(y_0(0), y_1(0) , y_2(0))=(y_0,
y_1, y_2)$ as well as $(\bar Y_0(0), \bar Y_1(0) , \bar Y_2(0)) \to
(y_0, y_1, y_2)$ as $n\to \infty$, such that $\sum_jy_j=1$ (the
proportions sum up to 1) and $\sum_jjy_j/2=i(0)>0$ (there is a
positive initial fraction of infectives).

Ball \cite{Ball99} also derives a threshold parameter
$R_*=(\lambda_G/\gamma)(1+\lambda_L/\gamma)$ such that the
deterministic epidemic will go extinct (i.e.\ $y_0(t)\to 1$ as $t\to
\infty$) if $R_*\le 1$, whereas it will reach an endemic level if
$R_*>1$. In the latter case the endemic level $(\hat y_0,\hat y_1,
\hat y_2)$ is given by Ball \cite{Ball99}, Theorem 3.1. This gives
that the endemic level of infectives $\hat i=\sum_jj\hat y_j/2$
equals
\begin{equation}
\hat i=  \frac{2 \lambda_L \gamma}{\lambda_G (
\sqrt{(\lambda_L+\lambda_G)^2 + 4 \lambda_L\gamma} - (\lambda_L
+\lambda_G))} - \frac{\gamma + \lambda_L}{\lambda_G}. \label{i-hat}
\end{equation}


\section{The time to extinction for the SIS household epidemic}

In the present section we derive a rough approximation for the time
to extinction of the SIS household epidemic model defined in Section
\ref{SIS-def} applicable for all values of $\lambda_L$ and relying
on $n$ being large, and a more precise approximation suitable for
large values of $\lambda_L$. Before doing this we present the
results of Andersson and Djehiche \cite{AD98} for the time to
extinction of the homogeneous SIS epidemic model, i.e.\ the model
above but without having household structure corresponding to the
case where $\lambda_L=0$. Without loss of generality we shall fix
$\gamma =1$ which simply means that we measure time with the average
infectious period as the base time unit.

\subsection{The time to extinction for the homogeneous SIS model}
When there are no households ("the homogeneous case") the process is
described by $(S(t),I(t))$, the number of susceptible and the number
of infectives, but since $S(t)+I(t)=N$ is fixed it is enough to keep
track of $I(t)$ say. When $N$ is large and assuming
$R_*=\lambda_G>1$, $I(t)$ will fluctuate around the endemic level
$N\hat i=N(1-1/\lambda_G)$ but eventually make a big "excursion"
down to $I(t)=0$ when the epidemic goes extinct. Andersson and
Djehiche \cite{AD98} prove that, as $N$ tends to infinity, the time
$\tau^{(AD)}_N$ until this happens is exponentially distributed with
mean
\begin{equation} \label{mean_ext_homo}
E(\tau^{(AD)}_N)\sim
\sqrt{\frac{2\pi}{N}}\frac{\lambda_G}{(\lambda_G -1)^2} e^{N(\log
\lambda_G -1+1/\lambda_G)}.
\end{equation}
The idea behind the proof is that the process will make many
excursions from the endemic level eventually returning to the
endemic level. Finally it will make a big excursion down to the
absorbing state $I(t)=0$ when the epidemic goes extinct. It will
hence make a geometric number of returns to the endemic level before
going extinct, and in the proof they derive the expected duration of
an excursion which returns to the endemic level and the (small)
probability of making a big excursion to extinction.

\subsection{Approximation of the SIS household epidemic}\label{O-U-appr}

As mentioned in Section \ref{SIS-prop} it was shown by Ball
\cite{Ball99} that when $N$ is large the SIS household epidemic may
be approximated by a deterministic process. Using theory for Markov
population processes (Ethier and Kurtz \cite{EK86}) it is also
possible to show a weak convergence result. Our process has three
components, but because the total number of households
$\sum_{i=0}^2Y_i(t)=n (=N/2)$ is fixed it is really two dimensional.
This means it is enough to keep track of two components, e.g.\ $Y_0$
and $Y_1$. Since our prime interest is the total number of
infectives, it is natural to follow House and Keeling \cite{HK08}
and parameterise the model in terms of $I (t)$ and $Y_1 (t)$.
Suppose the process is started from the endemic level, i.e.\ $(I
(0),Y_1(0))=(N\hat{i}, N\hat y_1/2)$, and define the standardized
process by
\begin{equation}
\left(
\begin{array}{c} \tilde I^{(N)}(t)\\ \tilde Y^{(N)}_1(t)
\end{array}\right)
:= \sqrt{N}\left(
\begin{array}{c} \bar I(t) - \hat{i}\\ \bar Y_1(t)/2 - \hat y_1/2
\end{array}\right).
\end{equation}
Then it follows from Ethier and Kurtz \cite{EK86} that the vector
process $\tilde Z^{(N)}(t)$ with components $\tilde I^{(N)}(t)$ and
$\tilde Y^{(N)}_1(t)$ converges weakly to $\tilde Z$, where $\tilde
Z$ is an Ornstein-Uhlenbeck process. The properties of the
Ornstein-Uhlenbeck process stem from the system of differential
equations (\ref{det-diff}), where we now have assumed that
$\gamma=1$. These differential equations can in vector form be
written as $z'(t)=F(z(t))$, where $z(t)=(i (t), y_1(t)/2)$. Using
this notation, $\tilde Z$ is defined by the drift matrix
\begin{equation}
B=\partial F = \left( \begin{array}{cc} \lambda_G (1 - 2 i(t)) -1 &
\lambda_L
\\
\lambda_G (1-2i(t)) +1 - \lambda_G y_1(t)/2 & -1 - \lambda_L,
\end{array}\right), \label{B-def}
\end{equation}
and local covariance matrix
\begin{equation}
S= \left( \begin{array}{cc} \beta_1(t) + \beta_2(t) + \beta_3(t) +
\beta_4(t) &
 \beta_1(t) - \beta_2(t) -\beta_3(t) + \beta_4(t)
\\
 \beta_1(t) - \beta_2(t) -  \beta_3(t) + \beta_4(t) &
 \beta_1(t) + \beta_2(t) + \beta_3(t) + \beta_4(t). \label{S-def}
\end{array}\right),
\end{equation} where $\beta_1(t) = \lambda_G i(t) (1-i (t)- y_1(t)/2)$, $\beta_2(t)
=(\lambda_G i(t) y_1(t)/2 + \lambda_L y_1(t))$, $\beta_3 (t)=  2
(i(t)-y_1(t)/2)$ and $\beta_4 (t)= y_1(t)$ are the infinitesimal
transition rates of an infection within an household with no
infectives, an infection within an household with one infective, a
recovery within a household with two infectives and a recovery
within a household with one infective, respectively. Note that, if
$(i (0), y_1 (0)) = (\hat{i}, \hat{y}_1)$, then for all $t$, $(i
(t), y_1 (t)) = (\hat{i}, \hat{y}_1)$.

This Ornstein-Uhlenbeck has a Gaussian stationary distribution with
mean-zero and covariance matrix $\Sigma$ defined by
\begin{equation}
B\Sigma + \Sigma B^T = -S\label{Sigma-def}.
\end{equation}
This means that, for large $t$, our original process $Z(t)$ will,
conditional upon not having gone extinct, be approximately normal
with mean vector $(N\hat i, N\hat y_1/2)$ and covariance matrix
$N\Sigma$.

Solving $\Sigma$ is straightforward. However, the expression for
$\Sigma$ in terms of $\lambda_G$ and $\lambda_L$ is not insightful
as can be seen by the expression for $\Sigma_{11}$, the variance of
the total proportion of infectives, given by,
\begin{eqnarray} \label{sigma-formula}
\Sigma_{11} &=& \frac{(\varsigma_2 - \varsigma_1)}{ ((\lambda_L +
\lambda_G)^2 + 4 \lambda_L) \varsigma_3 - \varsigma_4} \hat{i},
\end{eqnarray}
where
\begin{eqnarray*}
\varsigma_1 & =& 6\lambda_L^5+22 \lambda_G \lambda_L^4+36
\lambda_L^4+52 \lambda_L^3+30 \lambda_G^2 \lambda_L^3+76 \lambda_G
\lambda_L^3+18 \lambda_G^3 \lambda_L^2+52 \lambda_L^2 \lambda_G^2+8
\lambda_L^2 \\ && +42 \lambda_L^2 \lambda_G+10 \lambda_G^2
\lambda_L+4 \lambda_G^4 \lambda_L+14 \lambda_G^3 \lambda_L+2
\lambda_G^4
\\
\varsigma_2 &=& 2\sqrt{(\lambda_L + \lambda_G)^2+4\lambda_L}(3
\lambda_L^4+8 \lambda_G \lambda_L^3+12 \lambda_L^3+7 \lambda_L^2
\lambda_G^2+16 \lambda_L^2 \lambda_G+8 \lambda_L^2+2 \lambda_G^3
\lambda_L  \\ && +3 \lambda_L \lambda_G+6 \lambda_G^2
\lambda_L+\lambda_G^3)) \\
\varsigma_3 &=& (5 \lambda_L^4+12 \lambda_G \lambda_L^3+23
\lambda_L^3+9 \lambda_L^2 \lambda_G^2+23 \lambda_L^2+31 \lambda_L^2
\lambda_G+13 \lambda_G^2 \lambda_L+2 \lambda_L+2 \lambda_G^3
\lambda_L \\ && +14 \lambda_L \lambda_G+2 \lambda_G^3+2
\lambda_G^2) \\
\varsigma_4 &=& \sqrt{(\lambda_L + \lambda_G)^2+4\lambda_L} (15
\lambda_G^3 \lambda_L+21 \lambda_G^2 \lambda_L^3+48 \lambda_L^2
\lambda_G^2+59 \lambda_L^3+33 \lambda_L^4 \\ && +5 \lambda_L^5+22
\lambda_L^2 +2 \lambda_G^3+2 \lambda_G^4+68 \lambda_G \lambda_L^3+17
\lambda_G \lambda_L^4+6 \lambda_L \lambda_G \\ && +20 \lambda_G^2
\lambda_L+59 \lambda_L^2 \lambda_G+11 \lambda_G^3 \lambda_L^2+2
\lambda_G^4 \lambda_L)).
\end{eqnarray*}
Similar expressions exist for the other components of $\Sigma$. For
$k >2$, the corresponding expression to (\ref{sigma-formula}) is
even more unwieldily. However, studying the behaviour of
(\ref{sigma-formula}) in the limits as $\lambda_L \downarrow 0$ and
$\lambda_L \rightarrow \infty$ is informative.

Let $\sigma^2_i (\hat{i}, \lambda_L) = \Sigma_{1,1}$, the variance
for the total proportion of infectives in equilibrium, explicitly
stating the dependence upon $\hat{i}$ and $\lambda_L$. (Note that
$\lambda_G$ can be expressed as a function of $\hat{i}$ and
$\lambda_L$.) Then for fixed $\hat{i}$, $\sigma^2_i (\hat{i}, 0) =
\hat{i}^{-1}$ and $\lim_{\lambda_L \rightarrow \infty} \sigma^2_i
(\hat{i}, \lambda_L) = 2 \hat{i}^{-1}$. That is, for large
$\lambda_L$ the variance of the proportion of infectives is
approximately twice the corresponding variance in the homogeneous
case. There is a simple explanation for this. In the homogeneous
case, $\lambda_L =0$, we have $N=2n$ individuals. On the other hand,
when $\lambda_L \rightarrow \infty$, the two members of a household
are effectively paired together, either both susceptible or both
infectious. Thus the population in effect consists of $n$ paired
individuals resulting in the variance doubling.

This leads onto the question of, whether or not, for fixed
$\hat{i}$, $\sigma^2_i (\hat{i}, \lambda_L)$ is monotonically
increasing in $\lambda_L$. Plots of $\sigma^2_i (\hat{i},
\lambda_L)$, fixing $\hat{i}$ and varying $\lambda_L$, suggest that
this is the case if $\hat{i} \geq 0.5$, whilst, $\sigma^2_i
(\hat{i}, \lambda_L)$, is initially decreasing if $\hat{i} < 0.5$.
This is partially confirmed by studying $\left. \frac{\partial
\;}{\partial \lambda_L} \sigma^2_i (\hat{i}, \lambda_L)
\right|_{\lambda_L =0}$ which is positive, equal to 0 and negative
when $\hat{i} >0.5$, $\hat{i} =0.5$ and $\hat{i} < 0.5$,
respectively. Note that for general choices of $k$, $\sigma_i^2
(\hat{i},0) = \hat{i}^{-1}$ and $\lim_{\lambda_L \rightarrow \infty}
\sigma_i^2 (\hat{i},\lambda_L) = k \hat{i}^{-1}$.

If instead we consider fixed $R_\ast = \lambda_G (1 + \lambda_L)$,
$\Sigma_{11}$ is increasing as $\lambda_L$ increase, for all $R_\ast
>1$. Also $\hat{i}$ is maximised at the extremes $\lambda_L =0$ and
$\lambda_L \rightarrow \infty$, where $\hat{i} = 1 - 1/R_\ast$.

The above analysis gives a good description of the endemic level, we
now turn to the question of time to extinction and look to see how
the above Ornstein-Uhlenbeck approximation can be used to assist in
estimating this quantity.

\subsection{The time to extinction for the SIS household epidemic}\label{T-approx}

It seems hard to derive an explicit result corresponding to that of
Andersson and Djehiche \cite{AD98} for the household epidemic, the
reason being that the process need not return to the endemic level
$(N\hat i, N\hat{y}_1/2)$, and also because the trajectory down to
extinction is not unique. It seems possible to derive a large
deviation result but not to obtain a useful explicit expression for
the time to extinction. Instead we have taken the approach first
used by N{\aa}sell \cite{Nas99} who applies it to the homogeneous
SIR epidemic with demography, also used in Andersson and Britton
\cite{AB00}. We now present this approximation.

Let $Q$ denote the quasi-stationary distribution of the  SIS
household epidemic, i.e.\ $Q$ is the stationary distribution of the
process $Y(t)=(Y_0(t),Y_1(t),Y_2(t))$ conditioned on not having gone
extinct. Hence,
\[
(Y(t)| Y(0)\sim Q, I(t)>0) \sim Q.
\]
Starting in the quasi-stationary distribution it follows,  because
of the memoryless property, that the time to extinction $T_Q=\inf
\{t; I(t)=0| Y(0)\sim Q\}$ is exponentially distributed with
intensity equal to the probability of being one step away from
extinction multiplied by the rate of moving into absorption. If we
denote the quasi-stationary distribution $Q=\{
q_{i,y_1}\}=P_Q(I(t)=i, Y_1(t)=y_1)$ we hence have that
\begin{equation}
T_Q\sim \mathrm{Exp} (\gamma q_{1,1}) = \mathrm{Exp} (q_{1,1}),
\end{equation} since we have taken $\gamma =1$.
Note that this is an exact result for any $N$. However, it remains
to derive $q_{1,1}$. Since when $I(t)=1$ we have by necessity that
$Y_1(t)=1$ (and $Y_0(t)=n-1$), it is enough to look at the marginal
distribution $\{q_i\}$ of the number of infectives. An approximation
for $q_1$ is given by the normal approximation of the stationary
Gaussian distribution derived in Section \ref{O-U-appr}. This
approximation consists of computing the normal density for $I(t)$
(which is approximately normal with mean $N\hat i$ and variance
$N\sigma_i^2 (\hat{i}, \lambda_L)$) at the point $1$ and
conditioning on that $I(t)>0$. This approximation is of course
better in central parts of the distribution (around $I(t)\approx
N\hat i$) but in the absence of a better approximation we use it
also in the tail $I(t)=1$.

Using the expression for the endemic level $(\hat i, \hat y_1)$ and
the variance $\sigma_i^2 (\hat{i}, \lambda_L)$ we get the following
approximation for $q_1$
\begin{eqnarray}
q_1 & \approx & P(I(t)=1|I(t)>0, t\ \mbox{large}) \nonumber \\
&\approx & \frac{1}{\sqrt{2 \pi N} \sigma_i (\hat{i}, \lambda_L)}
\exp \left( -\frac{1}{2} \times \frac{(1 - N \hat{i})^2}{N
\sigma_i^2 (\hat{i}, \lambda_L)} \right)
 \nonumber \\
& \approx& \frac{1}{\sqrt{2 \pi N} \sigma_i (\hat{i}, \lambda_L)}
\exp \left( - \frac{N \hat{i}^2}{2 \sigma_i^2 (\hat{i}, \lambda_L)}
\right) ,\label{q_1}
\end{eqnarray}
giving our first approximation for the time to extinction of the SIS
household epidemic
\begin{equation}
E(T_Q)=\frac{1}{q_1}\approx \sqrt{2 \pi N}  \sigma_i (\hat{i},
\lambda_L) \exp \left(\frac{N \hat{i}^2}{2 \sigma_i^2 (\hat{i},
\lambda_L)} \right) .\label{E(T_Q)}
\end{equation}

The dominant term in our first approximation (\ref{E(T_Q)}) is $\exp
\left(\frac{N}{2} \times \frac{\hat{i}^2}{\sigma^2_i (\hat{i},
\lambda_L)} \right)$. Thus for large $N$ and a fixed value of
$\hat{i}$, the time to extinction is determined by $\sigma^2_i
(\hat{i}, \lambda_L)$ with the smaller the value of $\sigma^2_i
(\hat{i}, \lambda_L)$, the longer the time to extinction is expected
to be. From the study of  $\sigma^2_i (\hat{i}, \lambda_L)$ our
first approximation hence suggests that $E(T_Q)$ is monotonically
decreasing in $\lambda_L$ if $\hat{i}$ is fixed and larger than 0.5.
If on the other hand we fix $\hat{i}<0.5$, $E(T_Q)$ first increases
and then decreases with $\lambda_L$, and $E(T_Q)$ hence has a
(local) maximum for a small but positive $\lambda_L$. It was noted
in Doering {\it et al.}~\cite{Doer05}, that for the homogeneously
mixing case the Ornstein-Uhlenbeck approximation is only reasonable
for $R_\ast = 1 + C/N^{\frac{1}{3}}$, for $C > 0$ and this is also
likely to be the case for the household model. However,
(\ref{E(T_Q)}) still proves to be useful in gaining an understanding
of how $\lambda_L$ affects the extinction time.

We now consider the case where $\lambda_L$ is large and derive a
second more precise approximation of $E(T_Q)$. As noted in Section
\ref{O-U-appr}, for large values of $\lambda_L$, it is unlikely that
there will be just one infective in a household. Consider the
infection of a susceptible household making one member of the
household infectious and assume for the moment that no further
global infections are made with the household. Then after a waiting
time of $\mathrm{Exp} (1+ \lambda_L)$ either the second member of
the household is infected (with probability $\lambda_L/(1+
\lambda_L)$) or the infective recovers (with probability
$1/(1+\lambda_L)$). If the second member of the household becomes
infected the time until one of these individual recovers is
$\mathrm{Exp} (2)$. This recovery will shortly be followed by an
infection (with probability $\lambda_L/(1+ \lambda_L)$) or a second
recovery (with probability $1/(1+\lambda_L)$). Let $A_1, A_2,
\ldots$ be independent and identically distributed according to $A
\sim \mathrm{Exp} (2)$ and let  $B_0, B_1,  \ldots$ be independent
and identically distributed according to $B \sim \mathrm{Exp} (1 +
\lambda_L)$. Let $G \sim Geom (1/(1 + \lambda_L))$ with support on
$\mathbf{Z}^+$. Let $R$ be the time from infection of the household
until it recovers and $S$ be the sum of the total time infectious of
the two individuals in the household from global infection until
recovery. Then
\[ R = B_0 + \sum_{j=1}^G (A_j + B_j), \] where the sum is 0 if
$G=0$ and \[ S=  \sum_{j=0}^G B_j + 2 \sum_{j=1}^G A_j. \] Now
$\sum_{j=0}^G B_j \sim \mathrm{Exp} (1)$ and
\[  \sum_{j=1}^G A_j = \left\{ \begin{array}{ll} 0 & \mbox{with
probability } \frac{1}{1 + \lambda_L} \\ \mathrm{Exp} (2/(1+
\lambda_L)) & \mbox{with probability } \frac{\lambda_L}{1 +
\lambda_L}.
\end{array} \right.
\]
Consequently, for large $\lambda_L$, $R \approx \tilde{R} \sim
\mathrm{Exp} (2/(1+ \lambda_L))$ and whilst both members of the
household are infectious the household is generating global
infections at the points of a homogeneous Poisson point process with
rate $2 \lambda_G$. Since $\lambda_L$ is large, it is highly
unlikely that a global infectious contact ($\lambda_G$ is
necessarily small) with a household with at least one infective in
it will be with a susceptible individual. Therefore for large
$\lambda_L$, the total number of infectious households
(approximately the total number of infectives divided by 2) can be
approximated by an SIS epidemic in a homogeneously mixing population
of size $N/2$ with infection rate $2 \lambda_G$ and recovery rate
$2/(1 + \lambda_L)$. From (\ref{mean_ext_homo}), this gives our
second approximation for the time to extinction
\begin{equation}
E(T_Q)\approx \frac{1 + \lambda_L}{2} \times
\sqrt{\frac{2\pi}{N/2}}\frac{R_\ast}{(R_\ast -1)^2}
e^{\frac{N}{2}(\log R_\ast -1+1/R_\ast)}, \label{mean_ext_homo2}
\end{equation} where
$R_\ast = \lambda_G (1 + \lambda_L)$. Thus the time to extinction
depends upon $\lambda_L$ and $N$. This suggests that the mean time
to extinction will approximately grow linearly in $\lambda_L$, for
fixed $\hat{i}$, as $\lambda_L \rightarrow \infty$. (For large
$\lambda_L$, $R_\ast \approx 1/(1 - \hat{i})$.)

This result can be extended to households where $k >2$. In general,
for large values of $\lambda_L$, the number of infectious households
can be approximated by a homogenously mixing SIS epidemic model with
infection rate $k \lambda_G$ and recovery rate $k \prod_{j=1}^{k-1}
\frac{k-1}{k-1 + \lambda_L (k-j)}$. For fixed $N$, the mean
extinction time will behave like $C_N \lambda_L^{k-1}$ as $\lambda_L
\rightarrow \infty$ for some $C_N >0$. Also
 $\lim_{\lambda_L \rightarrow \infty} \sigma^2_i (\hat{i},
\lambda_L) = k \hat{i}^{-1}$. Thus the effect of $\lambda_L$
(households) on mean time to extinction is more marked for larger
values of $k$. A similar result holds for unequal sized households.

There is an apparent contradiction between the two approximations
with the Ornstein-Uhlenbeck approximation (\ref{E(T_Q)}) stating
that  $E(T_Q)$ \emph{decreases} in $\lambda_L$ for large enough
$\lambda_L$ and the second approximation (\ref{mean_ext_homo2})
stating that $E(T_Q)$ \emph{increases} in $\lambda_L$ as
$\lambda_L\to\infty$. The explanation for this is that there are two
asymptotic regimes considered: $N \rightarrow \infty$
(Ornstein-Uhlenbeck approximation) and $\lambda_L \rightarrow
\infty$ (second approximation). For fixed population size, $N$,
there is a cross-over from the Ornstein-Uhlenbeck approximation to
the second approximation as $\lambda_L$ increases. Hence this
suggests that as $\lambda_L$ changes from 0 to $\infty$ and when
$\hat{i}<0.5$ (the most common situation) the mean time to
extinction should first increase, then decrease and, when moving
over to the second approximation, eventually start increasing again.
If we instead consider the situation where $N$ increases, the
transition between the two approximations occurs at increasing
values of $\lambda_L$. Thus as $N \rightarrow \infty$, the
Ornstein-Uhlenbeck approximation dominates for all values of
$\lambda_L$.

Equations (\ref{mean_ext_homo}) and (\ref{mean_ext_homo2}) give
approximate mean times to extinction in the cases where $\lambda_L$
is close to 0 and $\lambda_L$ is large, respectively. The question
remains of estimating the time to extinction for moderate values of
$\lambda_L$ since as noted by Doering {\it et al.}~\cite{Doer05} the
Ornstein-Uhlenbeck estimation of $q_1$ can be several orders of
magnitude too small even for moderate $N$. However, the
Ornstein-Uhlenbeck approximation can be used for a qualitative
assessment of how the mean time to extinction changes when departing
from the homogeneous case. That is, we can compare the
Ornstein-Uhlenbeck approximation for $q_1$ under the assumption of
homogeneously mixing and for moderate values of $\lambda_L$ for a
given value of $\hat{i}$. For example, for $\hat{i} < 0.5$, does the
mean extinction time increase for small values of $\lambda_L$ before
decreasing as suggested by the Ornstein-Uhlenbeck approximation?
Furthermore, what are the competing influences of $N$ and
$\lambda_L$ through the two approximations on the mean time to
extinction? These questions are addressed in the following section.

\section{Numerical examples and simulations}

For the numerical examples and simulations we focus attention upon
$\hat{i} =0.2$. In the homogeneous case this corresponds to
$\lambda_G =1.25$. Thus $\sigma^2 (0.2,0) = 5$ and $\lim_{\lambda_L
\rightarrow \infty} \sigma^2 (0.2, \lambda_L) =10$. A simulation
study involving $N=50$ and $N= 200$ and $\lambda_L =0,
0.1,0.2,\ldots, 1, 2, \ldots, 10$ was run, with for each set of
parameter values the mean time to extinction estimated from 10000
simulations starting at the endemic level (see Figure 1). For
$N=50$, the mean time to extinction was found to grow linearly in
$\lambda_L$. The correlation between $\lambda_L$ and mean time to
extinction was found to be $0.9992$.  The case $\lambda_L =50$ was
also tested and found to satisfy the trend found for smaller values
of $\lambda_L$.  We conclude that the Ornstein-Uhlenbeck process
(relying on $N$ to be large) is not applicable for this case, and
the second approximation is better for the whole range of
$\lambda_L$. However, as $N$ increased a different story emerged.
First, for very small $\lambda_L$ $E(T_Q)$ seems to increase
slightly and then decreases up until $\lambda_L\approx 2.5$. Hence
this part agrees with the behaviour suggested by the first
approximation. After this, i.e.\ for $\lambda_L>2.5$, $E(T_Q)$
starts growing close to linearly with $\lambda_L$ as suggested by
the second approximation (\ref{mean_ext_homo2}).
\begin{figure}
 \psfrag{lambda_L}[][]{{\footnotesize $\lambda_L$}} \epsfig{file=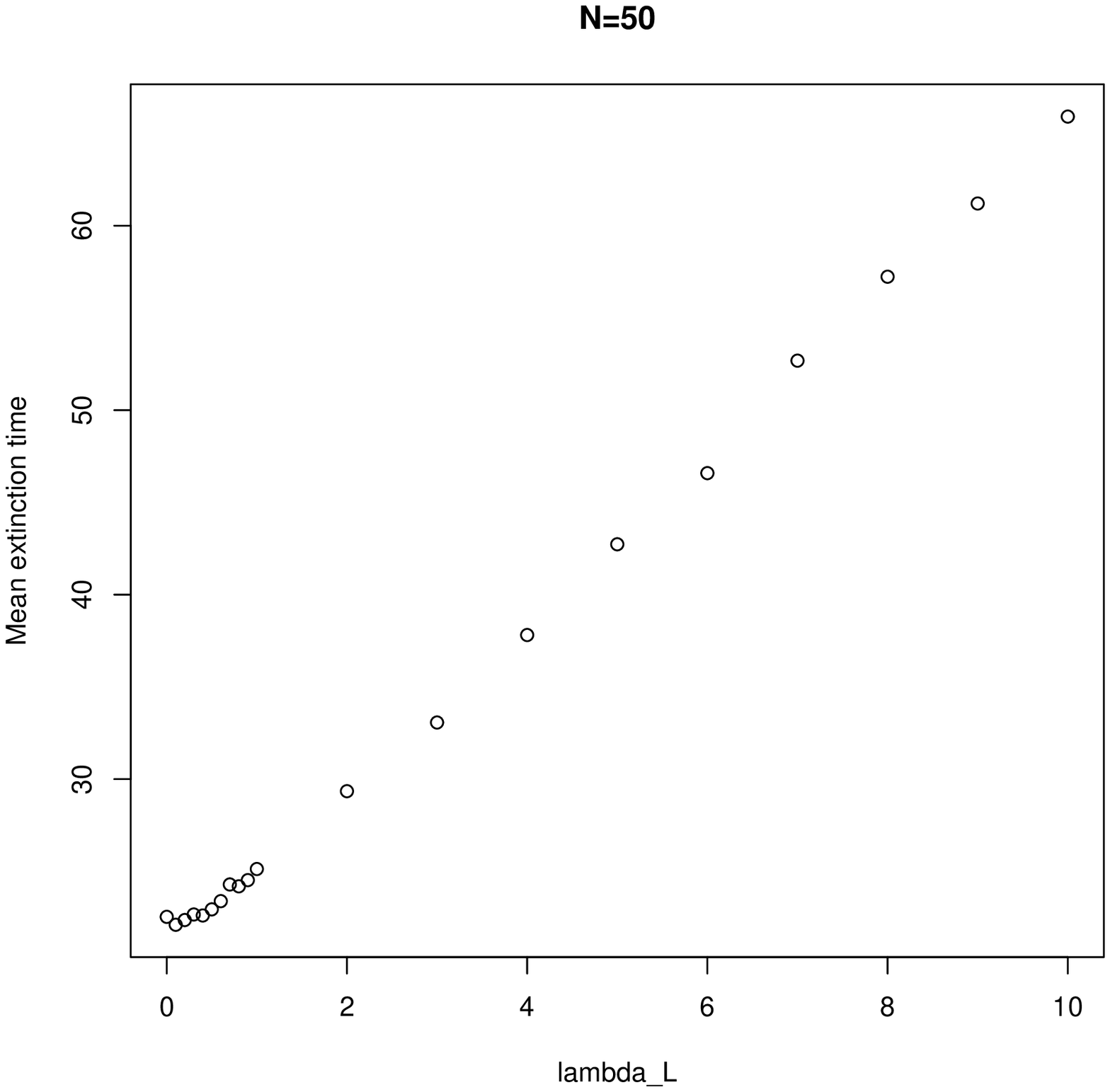,
height=6cm, width=6cm} \epsfig{file=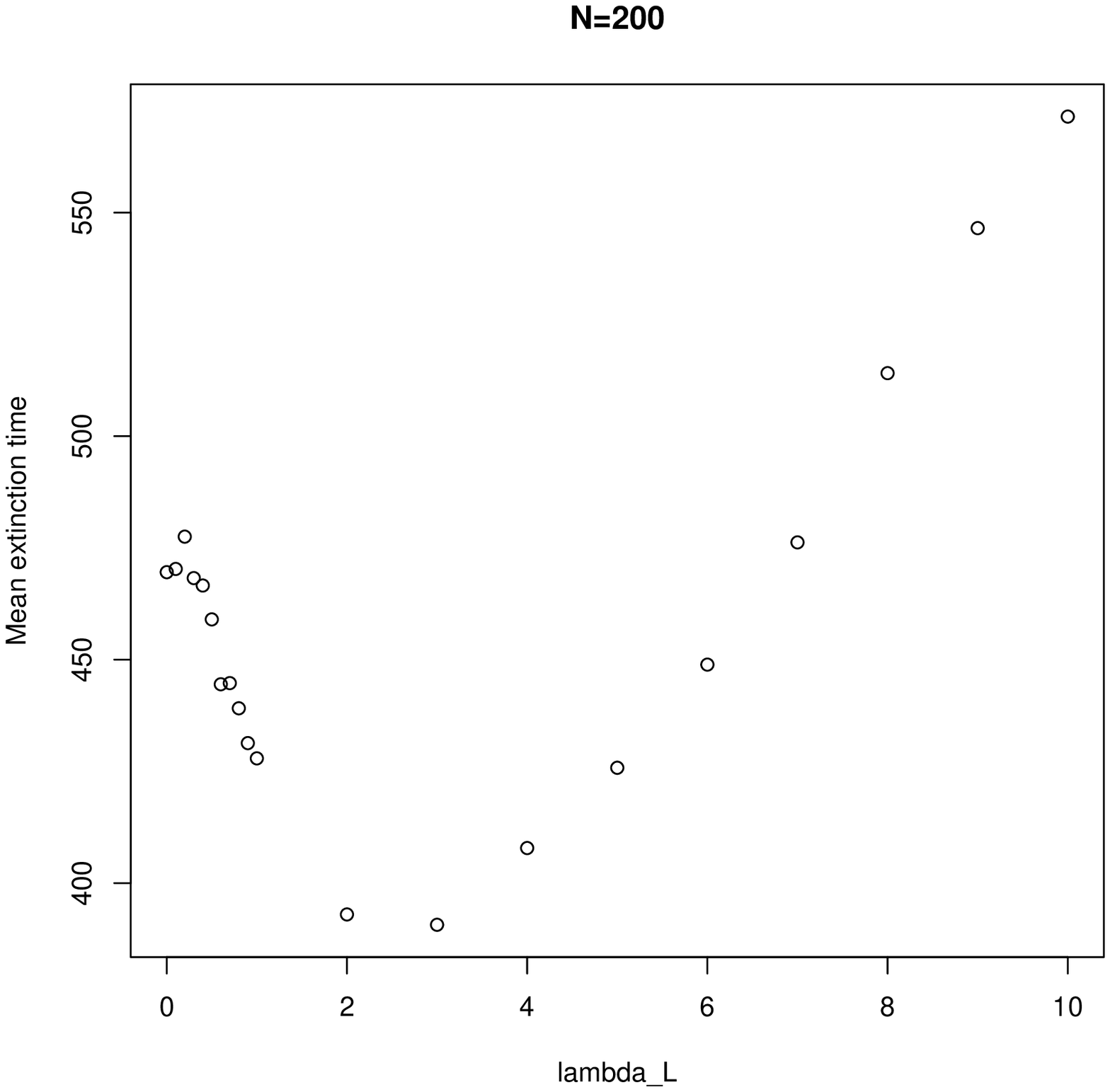, height=6cm,
width=6cm}
\\ {\it Figure 1: Mean extinction times from 10000 simulations: Left figure $N=50$ and right figure
$N=200$.} \end{figure}

Similar results can be obtained where all households are of size
$k>2$. In such cases, as $\lambda_L \rightarrow \infty$, the mean
extinction time for fixed $N$ and $\hat{i}$ increases polynomially
in $\lambda_L$. For moderate values of $\lambda_L$ and $\hat{i} <
0.5$, the mean time to extinction is maximized by small (but
positive) values of $\lambda_L$ as with households of size 2.  This
is illustrated in Figure 2 below where households of size $k=5$ are
considered, and where the mean extinction time has been simulated
for $N=50$ and $N=200$. The results are similar with those for
$k=2$, only more markedly. When $N=50$ the second approximation is
better for all $\lambda_L$ (suggesting polynomial growth of $E(T_Q)$
as a function of $\lambda_L$) whereas when $N=200$ the first
approximation (Ornstein-Uhlenbeck), suggesting that $E(T_Q)$ should
first increase and then decrease, works for small and moderate
$\lambda_L$, and after that, the second approximation suggesting
polynomial growth in $\lambda_L$ performs better.

\begin{figure}
 \psfrag{lambda_L}[][]{{\footnotesize $\lambda_L$}} \epsfig{file=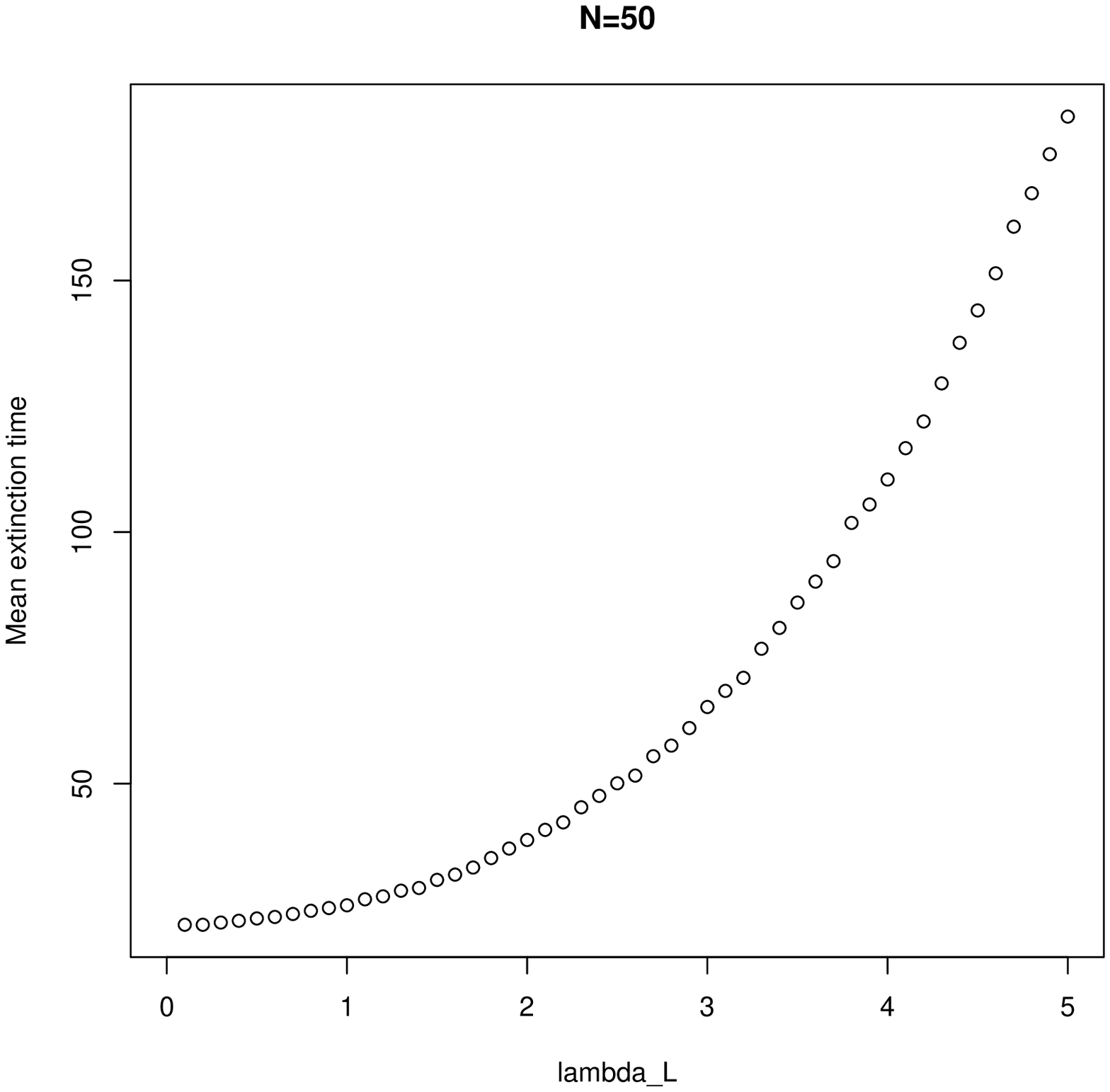,
height=6cm, width=6cm} \epsfig{file=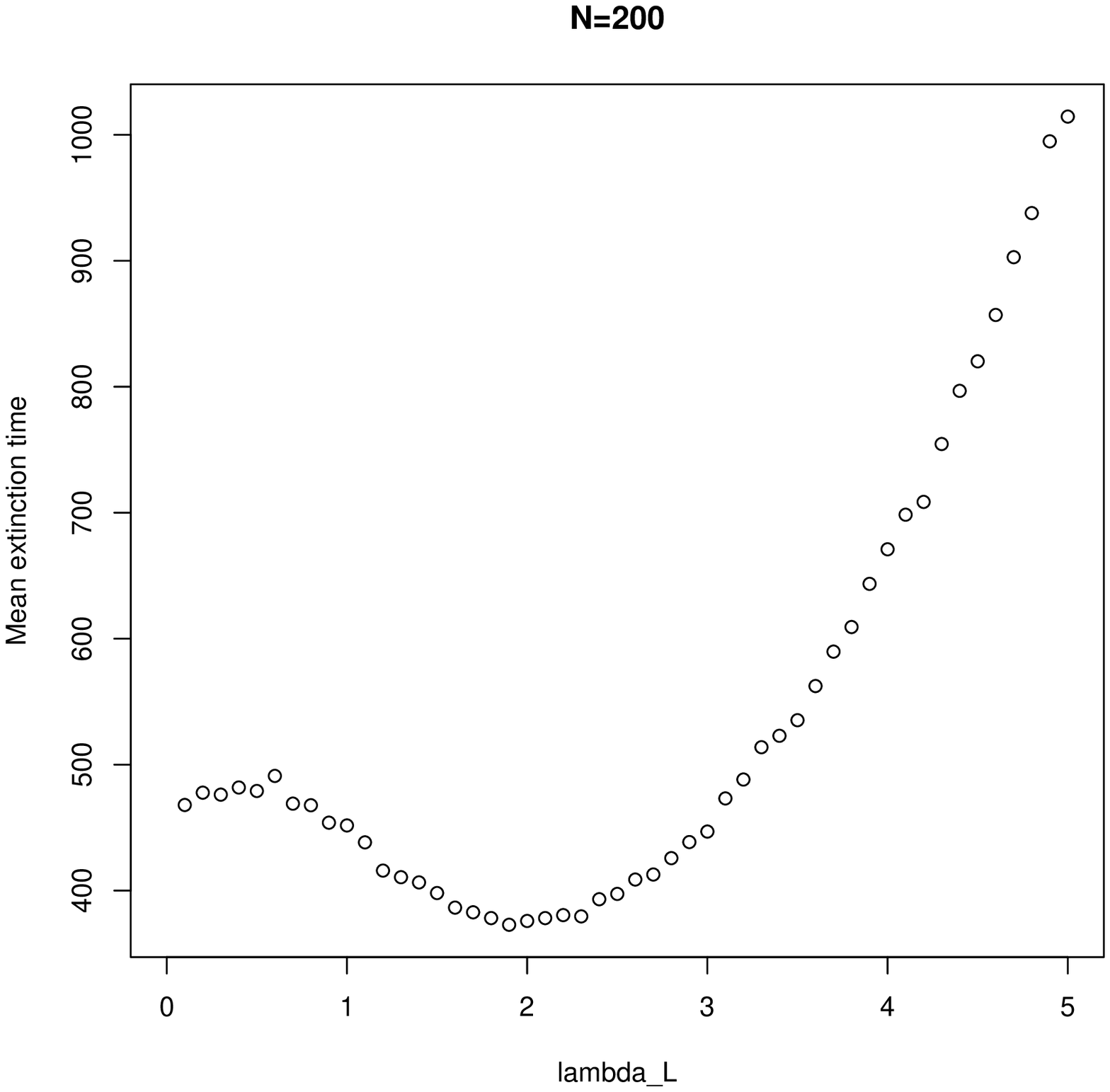, height=6cm,
width=6cm}
\\ {\it Figure 2: Mean extinction times from 10000 simulations in the case $k=5$: Left figure $N=50$ and right figure
$N=200$.} \end{figure}

In Figure 3, the ratio of the mean time to extinction with $N=200$
divided by  the mean time to extinction with $N=50$ is plotted. By
(\ref{E(T_Q)}), this ratio should be approximately,
\begin{eqnarray} \label{T_Q-ratio}
\frac{\sqrt{2 \pi \times 200} \sigma (\hat{i}, \lambda_L) \exp (200
\hat{i}^2/2 \sigma^2 (\hat{i}, \lambda_L))}{\sqrt{2 \pi \times 50}
\sigma (\hat{i}, \lambda_L) \exp (50 \hat{i}^2/2 \sigma^2 (\hat{i},
\lambda_L))} = 2 \exp (3/ \sigma^2 (\hat{i}, \lambda_L)).
\end{eqnarray}
Thus (\ref{T_Q-ratio}) states we should expect to see the ratio
between the extinction times initially increase before decreasing as
$\lambda_L$ goes from 0 to infinity, corresponding to $\sigma^2
(\hat{i}, \lambda_L)$ initially decreasing before increasing as
$\lambda_L \to \infty$. Figure 3 is consistent, at least
qualitatively, with the Ornstein-Uhlenbeck approximation. Similar
results are observed with $N=500$ with the Ornstein-Uhlenbeck
approximation `valid' for larger values of $\lambda_L$.
\begin{figure}
 \psfrag{lambda_L}[][]{{\footnotesize $\lambda_L$}} \epsfig{file=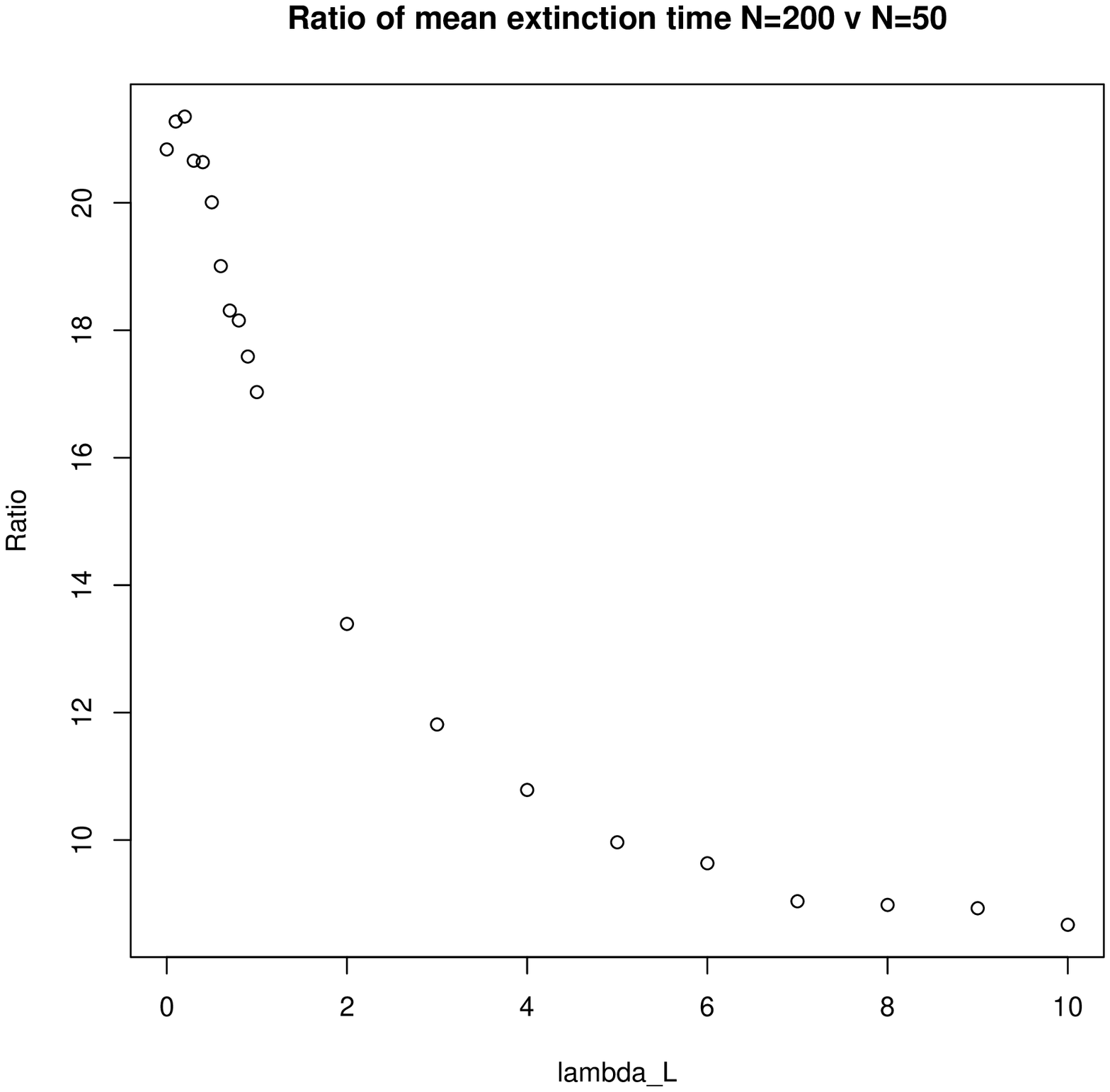,
height=6cm, width=6cm}  \epsfig{file=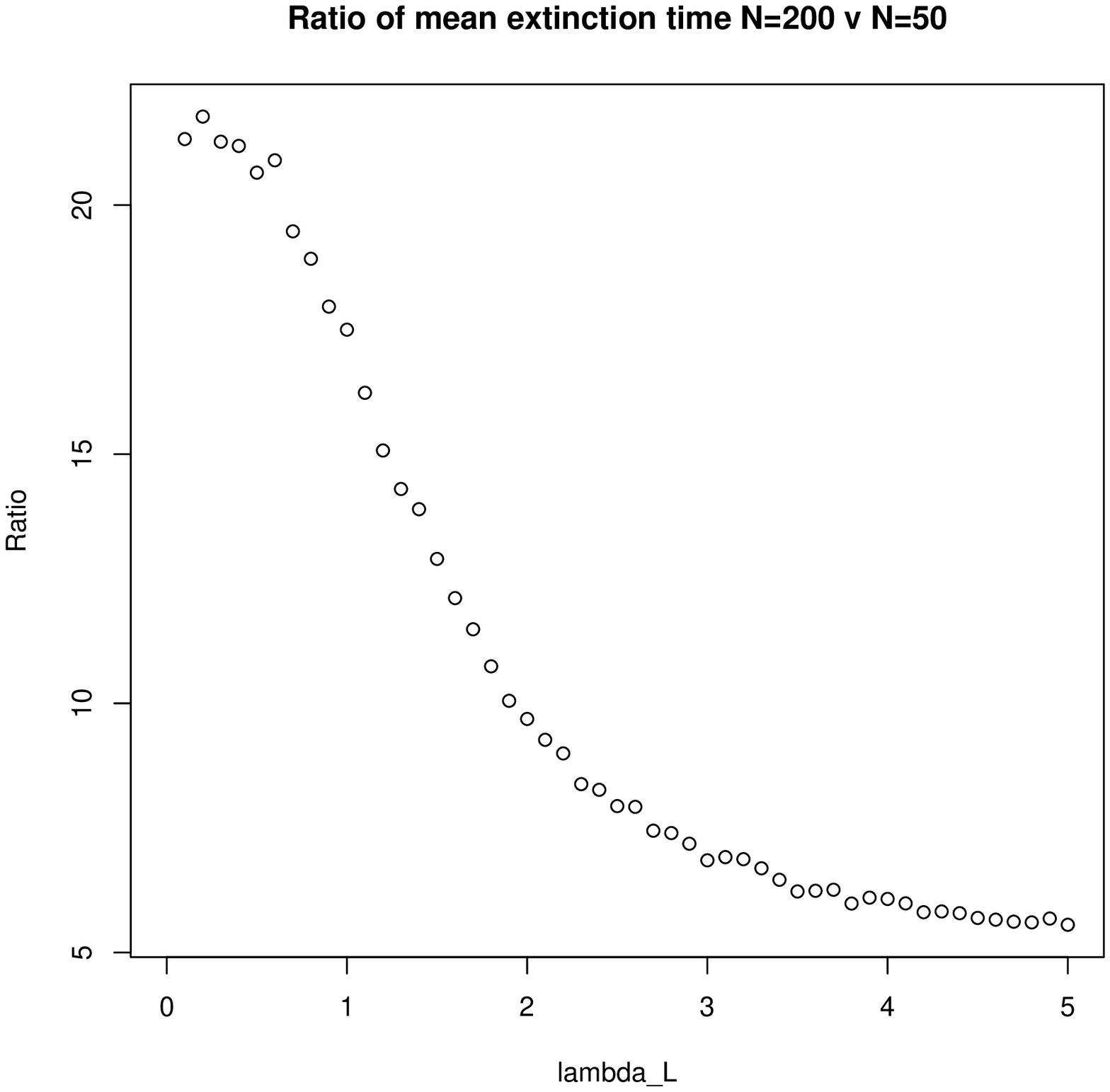, height=6cm,
width=6cm}
\\ {\it Figure 3: Ratio of mean extinction times from 10000 simulations for $N=200$ and $N=50$. Left figure $k=2$ and
right figure $k=5$} \end{figure}

For $\hat{i} > 0.5$, we would expect to see that the extinction time
decreases as $\lambda_L$ starts increasing from 0, in the limit as
$N \to \infty$. Analysis of this case is not presented for two
reasons. Firstly, in realistic situations $\hat{i}$ is likely to be
small and at the very least less than 0.5. Secondly, the mean
extinction time is very large even for small $N$. For example, for
$\hat{i} =0.8$ and $N=50$, in the homogeneous mixing case $E [
\tau_N^{(AD)}] = 4.18 \times 10^{16}$ compared with 22.55 when
$\hat{i} =0.2$.

\section{Discussion}

We present two approximations for obtaining the mean time to
extinction in an SIS household epidemic model. Approximating for
large values of $\lambda_L$ the SIS household epidemic model by a
homogeneously mixing epidemic model with the households treated as
individuals is informative and explains the trend observed for the
mean time to extinction in the simulation study as $\lambda_L$
increases. However, it is only really useful when $\lambda_L$ is
several orders of magnitude larger than $\lambda_G$. The most
interesting case is when $\lambda_G$ and $\lambda_L$ are of the same
order of magnitude. In that case we have resorted to the crude
Ornstein-Uhlenbeck approximation to gain an understanding in the
mean time to extinction. Although this approximation can severely
over-estimate the mean time to extinction it does perform very well
in giving a qualitative assessment of how the mean time to
extinction varies with $\lambda_L$.

As mentioned in the introduction, better approximations are in
principle available using large deviations, see for example, Shwartz
and Weiss \cite{SW95}. However, in practice it is difficult, if
indeed possible, to get an explicit expression for the large
deviations calculations.

The SIS epidemic model is the simplest epidemic model which exhibits
endemic behaviour. It would be interesting to extend the above
analysis to more realistic epidemic models with household structure.
A prime example would be an SIR epidemic model with demography
(births of susceptible individuals) extending the work of Andersson
and Britton \cite{AB00} to include household structure.

It would also be interesting to consider the mean time to extinction
in other SIS epidemic models with heterogeneous mixing of
individuals. Examples include the great circle model (Neal
\cite{Neal08}) and epidemics upon random graphs ({\it
e.g.}~Andersson \cite{And99}).

\begin{acknowledgements}
We would like to thank Frank Ball for showing us a copy of his
presentation, "Epidemics with two levels of mixing", presented at
DIMACS workshop: Stochasticity in Population and Disease Dynamics.
\end{acknowledgements}



\end{document}